\documentclass[showpacs,aps,prd,nofootinbib,floatfix,amsmath,amssymb]{revtex4}
\usepackage{graphicx}
\begin{document}

\makeatletter
\newbox\slashbox \setbox\slashbox=\hbox{$/$}
\newbox\Slashbox \setbox\Slashbox=\hbox{\large$/$}
\def\pFMslash#1{\setbox\@tempboxa=\hbox{$#1$}
  \@tempdima=0.5\wd\slashbox \advance\@tempdima 0.5\wd\@tempboxa
  \copy\slashbox \kern-\@tempdima \box\@tempboxa}
\def\pFMSlash#1{\setbox\@tempboxa=\hbox{$#1$}
  \@tempdima=0.5\wd\Slashbox \advance\@tempdima 0.5\wd\@tempboxa
  \copy\Slashbox \kern-\@tempdima \box\@tempboxa}
\def\FMslash{\protect\pFMslash}
\def\FMSlash{\protect\pFMSlash}
\def\miss#1{\ifmmode{/\mkern-11mu #1}\else{${/\mkern-11mu #1}$}\fi}
\makeatother

\title{Gauge invariance and radiative corrections in an extra dimensional theory}
\author{H. Novales--S\' anchez and J. J. Toscano}
\address{Facultad de Ciencias F\'{\i}sico Matem\'aticas,
Benem\'erita Universidad Aut\'onoma de Puebla, Apartado Postal
1152, Puebla, Puebla, M\'exico.}
\begin{abstract}
The gauge structure of the four dimensional effective theory originated in a pure five dimensional Yang-Mills theory compactified on the orbifold $S^1/Z_2$, is discussed on the basis of the BRST symmetry. If gauge parameters propagate in the bulk, the excited Kaluza-Klein (KK) modes are gauge fields and the four dimensional theory is gauge invariant only if the compactification is carried out by using curvatures as fundamental objects. The four dimensional theory is governed by two types of gauge transformations, one determined by the KK zero modes of the gauge parameters and the other by the excited ones. Within this context, a gauge-fixing procedure to quantize the KK modes that is covariant under the first type of gauge transformations is shown and the ghost sector induced by the gauge-fixing functions is presented. If the gauge parameters are confined to the usual four dimensional space--time, the known result in the literature is reproduced with some minor variants, although it is emphasized that the excited KK modes are not gauge fields, but matter fields transforming under the adjoint representation of $SU_4(N)$. A calculation of the one-loop contributions of the excited KK modes of the $SU_L(2)$ gauge group on the off-shell $W^+W^-V$, with $V$ a photon or a $Z$ boson, is exhibited. Such contributions are free of ultraviolet divergences and well--behaved at high energies.
\end{abstract}


\maketitle

\section{A general idea}
Standard Model (SM) extensions with TeV$^{-1}$-sized extra dimensions~\cite{antoniadis} involve at least three key scales, the first of which is the electroweak scale, at 246GeV, where spontaneous symmetry breaking takes place. Another main scale is the compactification one, whose size should be smaller than the current experimental sensitivity in order to make this sort of models phenomenologically consistent. Finally, the third scale is beyond the former ones, and there the physics should drastically change. Such asseverations suggest that any generalization of the SM to a given version with TeV$^{-1}$-sized extra dimensions is, from the commencement, an effective theory defined in a larger space--time manifold. A symptom of such a non--fundamental behavior is the fact that coupling constants in extra dimensional models have dimensions, which implies that these models are nonrenormalizable. When one compactifies the extra dimensions and integrates them, one obtains a four dimensionsal (4D) effective theory where the dynamic variables are the so-called KK modes, which are infinite in number. In the 4D space--time  the coupling constants are dimensionless, but the nonrenormalizable nature of the extra dimensional model is however inherited at the four dimensional level and manifests itself through the infinite KK towers. In connection with that, it has been proven~\cite{UED} that, in the case of one universal extra dimension, one-loop level effects are insensitive to the new physics scale and, in fact, the KK sums of the radiative corrections are convergent. By contrast, for two or more universal extra dimensions~\cite{UED} the loop effects are highly sensitive to the cut-off imposed by the fundamental scale, and the KK sums are rather divergent.

In this paper we shall start with a five dimensional (5D) Yang-Mills theory, and shall discuss a route to obtain a theory properly quantized at the 4D level. In order to reach such a goal, the fifth dimension shall be compactified on the orbifold $S^1/Z_2$ and then the extra coordinate shall be integrated in the action. By following this path, a 4D effective theory is obtained. Such a theory is constituted by the ordinary 4D Yang-Mills theory and other couplings among light gauge bosons and heavy ones. This 4D theory possesses an interesting gauge structure~\cite{NT}, which we shall discuss. We shall brefly comment on the employment of the Becchi-Rouet-Stora-Tyutin (BRST) formalism~\cite{BRST} to quantize this 4D effective theory. The quantum action so obtained can be utilized to perform radiative corrections on light Green functions. In particular, we shall exhibit the one-loop KK contributions to the $W^+W^-V$ vertex. Such contributions are well behaved~\cite{FMNRT} and  consistent with previous results reported in the literature. In fact, a comparison of such effects with the tree-level contributions of higher canonical dimension operators shows~\cite{FMNRT} that the former are dominant over the latter, as was already claimed in the case of one universal extra dimension~\cite{UED}.

\section{The gauge structure of the 4D effective theory}
We consider a 5D space-time manifold with a flat metric ${\rm diag}(1,-1,-1,-1,-1)$, where we define the $SU_5(N)$-invariant Lagrangian
\begin{equation}
\label{5dym}
{\cal L}_{5{\rm YM}}=-\frac{1}{4}{\cal F}^a_{MN}(x,y){\cal F}^{aMN}(x,y) ,
\end{equation}
with $x$ labeling the ordinary four dimensional space-time coordinates, while $y$ is the coordinate for the fifth dimension. Capital letters indices run over 0, 1, 2, 3, 5, and greek indices shall refer to the usual four-dimensional space-time. The 5D dynamic variables in Eq.(\ref{5dym}) are the vector bosons $({\cal A}^a_\mu (x,y),{\cal A}^a_5(x,y))$, which compose the 5D curvatures
\begin{equation}
{\cal F}^a_{MN}(x,y)=\partial_M {\cal A}^a_N(x,y)-\partial_N {\cal A}^a_M(x,y)+g_5f^{abc}{\cal A}^b_M(x,y){\cal A}^c_N(x,y).
\end{equation}
Here, $g_5$ is the 5D coupling constant, which has dimensions of $({\rm mass})^{-1/2}$. This Lagrangian is invariant under the set of gauge transformations
\begin{equation}
\delta {\cal A}^a_M={\cal D}^{ab}_M\alpha^b,
\end{equation}
which are defined by the gauge parameters $\alpha^b$. As we shall comment below, these parameters are crucial objects when deriving the 4D effective Lagrangian, as they determine two essentially different scenarios~\cite{NT}. One of them occurs when the gauge parameters propagate in the fifth dimension, whereas the other is defined by gauge parameters that do not depend on the fifth dimension. If the extra dimension is compactified on the orbifold $S^1/Z_2$ with radius $R$ and the gauge parameters depend on the extra dimension, such parameters can be expanded in Fourier series as
\begin{equation}
\alpha^a(x,y)=\frac{1}{\sqrt{2\pi R}}\alpha^{(0)a}(x)+\sum_{n=1}^\infty\frac{1}{\sqrt{\pi R}}\alpha^{(n)a}(x)\cos \left( \frac{ny}{R} \right).
\label{gauparexp}
\end{equation}
In extra dimensional models, the zero-modes of the fields are identified with the light fields corresponding to the already known particles in the 4D space--time, while the excited KK modes are heavy new fields. By virtue of this, one should expect the zero-mode gauge parameters to define the \emph{standard gauge transformations} (SGT) of the 4D Yang-Mills theory. On the other hand, the excited KK gauge parameters should define a different set transformations that we call \emph{nonstandard gauge transformations} (NSGT). We shall see that under such circumstances all the zero-mode fields are gauge fields, and that some KK excited modes fields are also gauge fields. By contrast, if gauge parameters do not propagate in the extra dimension, these parameters cannot be Fourier--expanded, which in turn implies that they remain the same before and after compactification. This suggests that in four dimensions there must be only SGT.

As there are two scenarios, one should be able to derive two different 4D effective Lagrangians. In the case of gauge parameters propagating in the extra dimension the Fourier expansions must be performed over the 5D curvatures~\cite{NT},
\begin{eqnarray}
{\cal F}^a_{\mu \nu}(x,y)&=&\frac{1}{\sqrt{2\pi R}}{\cal F}^{(0)a}_{\mu \nu}(x)+\sum_{n=1}^\infty\frac{1}{\sqrt{\pi R}}{\cal F}^{(n)a}_{\mu \nu}(x)\cos \left( \frac{ny}{R} \right) ,
\label{curvexps1}
\\ \nonumber \\
{\cal F}^a_{\mu 5}(x,y)&=&\sum_{n=1}^\infty\frac{1}{\sqrt{\pi R}}{\cal F}^{(n)a}_{\mu 5}(x)\sin \left( \frac{ny}{R} \right).
\label{curvexps2}
\end{eqnarray}
This procedure contrasts with the path most commonly followed in the literature, where gauge fields are expanded instead~\cite{commint}. The lesson to learn is that expansion and integration of the extra dimension over covariant objects leads to 4D covariant objects~\cite{NT}. In the other scenario, where gauge parameters do not depend on the extra dimension, the objects to expand and integrate over are the gauge fields~\cite{NT}.

\subsection{First scenario: gauge parameters propagate in the extra dimension}
We define the 4D Lagrangian
\begin{equation}
{\cal L}_{4{\rm YM}}\equiv \int_0^{2\pi R} dy{\cal L}_{5YM},
\end{equation}
and insert the expansions shown in Eqs.(\ref{curvexps1},\ref{curvexps2}) into it,  which results in~\cite{NT}
\begin{equation}
\label{4deffL}
{\cal L}_{4{\rm YM}}=-\frac{1}{4}\left( {\cal F}^{(0)a}_{\mu \nu}{\cal F}^{(0)a\mu\nu}+{\cal F}^{(m)a}_{\mu\nu}{\cal F}^{(m)a\mu\nu}+2{\cal F}^{(m)a}_{\mu 5}{\cal F}^{(m)a\mu 5} \right),
\end{equation}
where repeated indices, including the mode ones, denote a sum. This convention shall be followed during the rest of the paper. This 4D Lagrangian has an elegant structure, for it is constituted exclusively by objects that resemble the Yang-Mills curvature. In fact, the precise form of the curvatures ${\cal F}^{(0)a}_{\mu \nu}$, ${\cal F}^{(m)a}_{\mu \nu}$ and ${\cal F}^{(m)a}_{\mu 5}$ can be straightforwardly determined~\cite{NT}. A crucial feature of the 4D Lagrangian, Eq.(\ref{4deffL}), is that it is invariant under the SGT~\cite{NT}
\begin{eqnarray}
\delta A^{(0)a}_\mu&=&{\cal D}^{(0)ab}_\mu \alpha^{(0)b},
\label{SGT1}
\\
\delta A^{(m)a}_\mu&=&gf^{abc}A^{(m)b}_{\mu}\alpha^{(0)c}
\label{SGT2}
\\
\delta A^{(m)a}_5&=&gf^{abc}A^{(m)b}_5\alpha^{(0)c},
\label{SGT3}
\end{eqnarray}
as well as under the NSGT~\cite{NT}
\begin{eqnarray}
\delta A^{(0)a}_\mu&=&gf^{abc}A^{(m)b}_\mu\alpha^{(m)c},
\label{NSGT1}
\\
\delta A^{(m)a}_\mu&=&{\cal D}^{(mn)ab}_\mu\alpha^{(n)b},
\label{NSGT2}
\\
\delta A^{(m)a}_5&=&{\cal D}^{(mn)ab}_5\alpha^{(nb)} ,
\label{NSGT3}
\end{eqnarray}
where ${\cal D}^{(0)ab}_\mu$ is the ordinary 4D Yang--Mills covariant derivative, ${\cal D}^{(mn)ab}_\mu$ is a sort of covariant derivative, given by~\cite{NT}
\begin{equation}
{\cal D}^{(mn)ab}_\mu=\delta^{mn}{\cal D}^{(0)ab}_\mu-gf^{abc}\Delta^{mrn}A^{(r)c}_\mu ,
\label{KKD}
\end{equation}
and ${\cal D}^{(mn)ab}_5$ is~\cite{NT} an object that does not involve derivatives. The factors $\Delta^{mrn}$ in the covariant drerivative Eq.(\ref{KKD}) are sums of Kronecker deltas and their precise form is not important for the present discussion. These sets of gauge transformations can be derived at least by means of three different methods~\cite{NT}: Fourier analysis, the Dirac's method~\cite{Dirac} toghether with the Castellani's formalism~\cite{Castellani}, and the BRST formalism~\cite{BRSTrev}.  Note that the zero--modes $A^{(0)a}_\mu$ are gauge fields under the SGT, but under the NSGT they change in a way that resembles a transformation in the adjoint representation, although in this case there is a mixing of KK modes. On the other hand, the excited modes $A^{(m)a}_\mu$ are matter fields under the SGT, but gauge fields under the NSGT. Finally, the scalar fields $A^{(n)a}_5$ are pseudo-Goldstone bosons, as they can be eliminated from the theory~\cite{NT} by performing a nonstandard gauge transformation with gauge parameters $\alpha^{(n)a}=(R/n)A^{(n)a}_5$. Concerning the Dirac's method, the only fields that generate constraints are $A^{(0)a}_\mu$ and $A^{(n)a}_\mu$, and such constraints are all first class, which implies that these fields are gauge fields, a feature that is consistent with the form of transformations (\ref{SGT1}--\ref{NSGT3}). It is worth emphasizing that the 4D curvatures ${\cal F}^{(0)a}_{\mu\nu}$, ${\cal F}^{(m)a}_{\mu\nu}$ and ${\cal F}^{(m)a}_{\mu 5}$ transform covariantly~\cite{NT} under the SGT and the NSGT.

\subsection{Second scenario: gauge parameters do not propagate in the extra dimension}
If gauge parameters are not allowed to propagate in the extra dimension, they remain the same after compactification and integration of the extra dimension, and instead of an infinite number of them at the 4D level, there are only $N^2-1$ gauge parameters in four dimensions. With this in mind, one should only expect SGT-invariance after the fifth dimension is compactified and intagrated. In this case, the objects to expand in Fourier series are the gauge fields instead of the curvatures~\cite{NT}. After compactifying and integrating the fifth dimension, one obtains the 4D Lagrangian~\cite{NT}
\begin{equation}
\hat{{\cal L}}_{4{\rm YM}}\equiv \int_0^{2\pi R}dy {\cal L}_{5{\rm YM}}={\cal L}_{4{\rm YM}}+\Delta {\cal L} .
\end{equation}
Here, the ${\cal L}_{4{\rm YM}}$ Lagrangian is the one exhibited in Eq.(\ref{4deffL}), which was obtained within the context of the first scenario. The term $\Delta {\cal L}$ is hence a difference between the Lagrangians derived in the two scenarios, and it is given by~\cite{NT}
\begin{equation}
\Delta {\cal L}=-\frac{1}{4}g^2f^{abc}f^{ade}\left( \Delta^{rnpq}A^{(n)c}_\nu A^{(q)e\nu}-\Delta '^{rnpq}A^{(n)c}_5A^{(q)e}_5 \right) A^{(r)b}_\mu A^{(p)d\mu} ,
\end{equation}
where the factors $\Delta^{rnpq}$ and $\Delta '^{rnpq}$ are sums and differences of Kronecker deltas. The $\hat{\cal L}_{4{\rm YM}}$ Lagrangian is invariant under the SGT, Eqs.(\ref{SGT1},\ref{SGT2},\ref{SGT3}), but not under the NSGT, Eqs.(\ref{NSGT1},\ref{NSGT2},\ref{NSGT3}), consistently with the fact that there are no KK excitations for the gauge parameters that define the latter sort of gauge transformations. In this scenario the nature of the fields involved in the 4D theory is quite different from the one found in the first scenario, as the zero-mode fields, $A^{(0)a}_\mu$, are still gauge fields, but the vector KK excited modes, $A^{(m)a}_\mu$, are matter fields and are not gauge fields. An important difference among the two scenarios is that, in this second context, the scalar KK fields, $A^{(n)a}_5$, are not Pseudo-Goldstone bosons, but matter fields, which cannot be removed from the theory~\cite{NT}. The last comment concerning this scenario is that the Dirac's method leads~\cite{NT}, under these circumstances, to first class as well as second class constraints, suggesting that there are matter fields, which is in agreement with the above discussion.

\section{Quantization}
Through this section, the quantization of the 4D effective theory, in the context of the first scenario, will be briefly discussed. Such a quantization, which is based on the BRST formalism~\cite{BRSTrev}, leads~\cite{NT} to the most general Faddeev-Popov ghost sector. An interesting feature of this quantization scheme is that the quantization of the zero-modes and the excited ones can be performed independently of each other~\cite{NT}. As we are mainly interested in the one-loop contributions to light Green functions with KK excited modes, exclusively, inside the loops~\cite{FMNRT}, we shall quantize only the KK excitations, leaving gauge invariance with respect to the SGT, but removing the degeneracy associated with the NSGT. This does not mean that the invariance under the SGT cannot be eliminated. In fact, the gauge with respect to such transformations can be fixed as in the ordinary Yang-Mills theory or by a nonstandard quantization scheme, such as the background field method~\cite{BFM}.

According to the BRST formalism~\cite{BRSTrev}, one starts with a configuration space that comprehends the ghost fields from the beginning. Such ghost fields coincide with the gauge parameters, although they have opposite statistics. As a next step, the configuration space is enriched by the introduction of an antifield per each field of the theory. Also a set of auxiliary fields is inserted. In this larger configuration space a symplectic structure, called the \emph{antibracket}, is defined and the master equation imposed. Then a proper solution to the master equation must be obtained. Particularly, the solution to the master equation in the case of the ordinary 4D Yang-Mills theory is already known, and its generalization to the 5D version is straightforward. The obtainment of the proper solution for the 4D effective theory can be accomplished~\cite{NT} by compactifying the extra dimension and performing a Fourier analysis on the 5D proper solution. As mentioned above, one can introduce a gauge-fixing procedure that maintains gauge invariance under the SGT. Such a procedure depends crucially on the employing of an appropriate set of gauge-fixing functions, which shall appear later in the quantized version of the Lagrangian and shall impact directly on the structure of the Faddeev-Popov ghost term. The following SGT--covariant set of gauge-fixing functions is introduced~\cite{NT}:
\begin{equation}
\label{gffunc}
f^{(m)a}={\cal D}^{(0)ba}_\mu A^{(m)b\mu}-\xi\frac{m}{R}A^{(m)a}_5.
\end{equation}
These functions are rather similar to other introduced~\cite{mRTT} some years ago, in the context of the so-called \emph{331 model}~\cite{331}. Once the proper solution to the master equation of the effective 4D theory is known, one can derive~\cite{NT} the gauge-fixed action and consequently the effective Lagrangian
\begin{equation}
\label{effq}
{\cal L}_{{\rm eff}}={\cal L}_{{\rm 4YM}}+{\cal L}_{{\rm GF}}+{\cal L}_{{\rm FPG}} .
\end{equation}
This expression comprehends some very important results, such as the Lagrangian ${\cal L}_{4{\rm YM}}$, which was shown in Eq.(\ref{4deffL}). The second term is the gauge-fixing part, ${\cal L}_{{\rm GF}}$, which is consituted by the gauge-fixing functions, Eq.(\ref{gffunc}), and is given by
\begin{equation}
{\cal L}_{{\rm GF}}=-\frac{1}{2\xi}f^{(m)a}f^{(m)a}.
\end{equation}
where $\xi$ is the gauge parameter. The last part, ${\cal L}_{{\rm FPG}}$, is the most general Faddeev-Popov ghost term, as it includes up to quartic ghost couplings. Its precise expression is~\cite{NT}
\begin{eqnarray}
{\cal L}_{{\rm FPG}}&=&\bar{C}^{(m)c}\left(\frac{\partial f^{(m)c}}{\partial A^{(n)a}_\mu}{\cal D}^{(nr)ab \mu}- \frac{\partial f^{(m)c}}{\partial A^{(n)a}_5}{\cal D}^{(nr)ab}_5 \right)C^{(r)b}-\frac{1}{\xi}gf^{abc}\Delta^{mrn}f^{(m)a}\bar{C}^{(r)b}C^{(n)c} 
\nonumber \\ \nonumber \\ &&
+\frac{1}{2}g^2f^{abc}f^{cde}\Delta^{mpq}\bar{C}^{(p)d}\bar{C}^{(q)e}\left(C^{(0)b}C^{(m)a}+C^{(0)a}C^{(m)b}+\Delta^{mrn}C^{(r)b}C^{(n)a}\,  \right) ,
\end{eqnarray}
with $C^{(n)b}$ and $\bar{C}^{(n)b}$ denoting the ghost and antighost fields, respectively. The effective Lagrangian, Eq.(\ref{effq}), contains all the couplings among light and heavy fields, which are crucial pieces to perform radiative corrections on light Green functions.

\section{Radiative corrections: The $W^+W^-V$ vertex}

The principal prediction of extra dimensional models is the existence of new particles, which correspond to the KK excited modes. If there is enough energy available at the Large Hadron Collider, the lightest of such new particles could be directly produced there. However, if no signals are observed in that collider, the International Linear Collider could detect the KK excited modes through their virtual effects on low energy observables. In connection with that, an interesting process is the $W^+W^-$ production, which could be studied at the one-loop level by inserting KK excited modes into the loops of the light $W^+W^-V$ vertex, with $V$ an off--shell photon or $Z$ boson. It is worth mentioning that the one--loop contributions involving KK excited modes, exclusively, to any light Green function are~\cite{NT} renormalizable, as, under such circumstances, the only divergencies that arise are those already present in the standard Yang-Mills theory and can therefore be~\cite{NT} absorbed by the parameters of the light theory.

Consider the one-loop correction to the $W^+W^-V$ vertex at the one-loop level. Suppose that the loops are formed by KK excited virtual fields, exclusively, and that $V$ is off--shell. The calculation of this vertex leads~\cite{FMNRT} to the well-known parametrization~\cite{WWV} of such an interaction, at the one-loop level, when only CP-even quantities are involved. In this case, there arise three form factors, given by~\cite{FMNRT}
\begin{eqnarray}
\nonumber
A&=&\frac{g^2}{96\pi^2(4x_W-1)^3}\sum_{n=1}^\infty\left( -24x_n(1-4x_W)^2B_0(1)+48x_W[5(x_W-1) x_W+4x_n(4x_W-1)B_0(2)
\right.
\\ \nonumber \\ \nonumber
&&
-6[-4x_W(6x_W^2-2x_W+3)+x_n(64x_W^2-4)+1] B_0(3)+36Q^2x_W[-4x_W^3+2x_W^2+x_W
\\ \nonumber \\
&&\left.
+ 4x_n(x_W-1)(4x_W-1)]C_0-4\{ 1-2x_W[x_W(20x_W-33)+9] \} \right) ,
\label{ffA}
\\ \nonumber \\ \nonumber
\Delta \kappa_V&=&\frac{g^2}{96\pi^2(4x_W-1)^3}\sum_{n=1}^\infty \left\{ 48x_n[B_0(1)-B_0(2)]+12x_W(26x_W+1)[B_0(2)-B_0(3)]
\right.
\\ \nonumber \\ \nonumber
&&
-384x_nx_W[B_0(1)-2B_0(2)+B_0(3)]+ 768x_nx_W^2[B_0(1)-3B_0(2)+2B_0(3)] 
\\ \nonumber \\
&&\left.
-72Q^2x_W(x_W^2-12x_nx_W+x_W+3x_n)C_0 +12x_W(4x_W-1)(8x_W+3) \right\} ,
\label{ffk}
\\ \nonumber \\ \nonumber
\Delta Q_V&=&\frac{g^2}{96\pi^2(4x_W-1)^3}\sum_{n=1}^\infty\left\{ -1536x_nx_W^3[B_0(1)-B_0(3)]+96(x_W-2)x_W^2(6x_W+1)[B_0(2)-B_0(3)] \right.
\nonumber \\ \nonumber && +768x_nx_W^2[B_0(1)+B_0(2)-2B_0(3)] -96x_nx_W[B_0(1)+2B_0(2)-3B_0(3)] \\ \nonumber \\ \nonumber &&
+144Q^2x_W[4x_W^4-4(4x_n+1)x_W^3+2(6x_n+1)x_W^2-6x_nx_W+x_n]C_0 
\\ \nonumber \\ &&\left.
-24x_W(4x_W-1)[2x_W(6x_W-1)+1] \right\} ,
\label{ffQ}
\end{eqnarray}
where $Q$ is the momentum of the off--shell neutral boson, $x_W\equiv m_W^2/Q^2$ and $x_n\equiv m_n^2/Q^2$, with $m_n=n/R$, $n=1, 2, 3,\ldots$. The following short--hand notation for the Passarino-Veltman scalar functions has been also introduced: $B_0(1)\equiv B_0(0,m_n^2,m_n^2)$, $B_0(2)\equiv B_0(m_W^2,m_n^2,m_n^2)$, $B_0(3)\equiv B_0(Q^2,m_n^,m_n^2)$ and $C_0\equiv C_0(m_W^2,m_W^2,Q^2,m_n^2,m_n^2m_n^2)$. It is worth emphasizing that these form factors were calculated by employing the Feynman-'t Hooft gauge, as it has shown to be an election that leads to well--behaved results in nonconventional quantization approaches such as the background field method~\cite{BFM}. The form factors $\Delta \kappa_V$ and $\Delta Q_V$ are interesting as they define the magnetic dipole moment and the electric quadrupole moment of the $W$ boson. It is a remarkable feature of the form factors (\ref{ffk},\ref{ffQ}) that they are free of UV divergencies~\cite{FMNRT}. Such an insensitivity to the cut--off scale is in agreement with the results reported in the context of universal extra dimensions~\cite{UED}. Altough these form factors are gauge dependent, it turns out that~\cite{FMNRT} they are well behaved at high energies and decouple in the limit of a very large compactification scale. For $0.5\textrm{TeV}<R^{-1}<1.0\textrm{TeV}$, both $\Delta \kappa_V$ and $\Delta Q_V$ range~\cite{FMNRT} from approximately $10^{-5}$ to $10^{-6}$, at best. The magnitude of the form factors in this range of energies is of the same order than those generated by new gauge bosons in the 331 model~\cite{mRTT}. For a size of the fifth dimension of $R^{-1}\sim 1\textrm{TeV}$, the one-loop contribution of the KK modes to these vertices is about one order of magnitude lower than the corresponding Standard Model radiative correction. If the compactification scale is taken as $R^{-1}\sim 300\textrm{TeV}$, as it was realized in the case of one universal extra dimension~\cite{UED}, the magnitudes of $\Delta \kappa_V$ and $\Delta Q_V$ is~\cite{FMNRT} of ${\cal O}(10^{-4})$, which should be at the reach of the International Linear Collider.

Consider the operator of higher canonical dimension
\begin{equation}
{\cal O}=\frac{g_5\alpha_W}{M_S^2}\frac{\epsilon^{abc}}{3!}{\cal A}^a_{\lambda \rho}{\cal A}^{a\rho\nu}{\cal A}^{c\lambda}_\nu ,
\end{equation}
where $\alpha_W$ is an unknown parameter that can be determined in terms of fundamental constants if the theory lying beyond the cut--off scale is known. Another ingredient of this operator is the cut--off scale, denoted by $M_S$. In connection with the form factors (\ref{ffA},\ref{ffk},\ref{ffQ}), this operator generates the term
\begin{equation}
\frac{g\alpha_W}{M_S^2}\frac{\epsilon^{abc}}{3!}A^{(0)a}_{\lambda \rho}A^{(0)b\rho \nu}A^{(0)c\lambda}_\nu ,
\end{equation}
which produces tree-level corrections to the $W^+W^-V$ vertex. The magnitude of such corrections for a compactification scale ranging from 0.5TeV to 1.0TeV is~\cite{FMNRT} lower by an order of magnitude than the contributions introduced by the one-loop corrections discussed in the last section.

\section{Conclusions}
The structure of the 4D effective Lagrangian that emerges from a 5D Yang-Mills theory after compactification and integration of the extra dimension was discussed. It was emphasized that the election concerning whether the 5D gauge parameters can propagate in the fifth dimension or not is a crucial step, as it defines two essentially different contexts. If gauge parameters depend on the extra dimension there is an infinite number of KK excitations, for such parameters, that determine an infinite number of gauge transformations, which can be divided into two types: the standard gauge transformations and the nonstandard gauge transformations. It was asseverated that, in order to obtain a Lagrangian that is invariant under both types of gauge transformations, the objects to Fourier-expand, after compactification, are the curvatures, as such a procedure generates 4D covariant objects after the integration of the fifth dimension. Contrastingly, if the 5D gauge parameters are not supposed to propagate in the fifth dimension, there are only SGT at the 4D level and the Fourier expansions, allowed by the compactification of the extra dimension, must be performed on the 5D gauge fields. The 4D effective Lagrangians defined by each of these approaches are quite different. The quantization of the 4D effective theory within the first context (5D gauge parameters propagate in the fifth dimension) was discussed on the basis of the BRST formalism, and a SGT--covariant gauge--fixing procedure  was introduced. The gauge--fixing functions defined within such an approach are similar to another scheme proposed some years ago for the so-called 331 model. The quantized Lagrangian, which includes a SGT--invariant gauge--fixing term and the most general ghost sector, was shown. From such an expression, the Lagrangian linking the light and heavy physics can be extracted, which is crucial to perform one-loop corrections to light Green's functions. It occurs that the one-loop level radiative corrections to light Green's functions are renormalizable, which is not necessarily true for the case of higher order corrections. The one-loop corrections to the $W^+W^-V$ vertex, involving only KK excitations inside the loops, were calculated and three form factors emerged. The ones that define the CP--even electromagnetic properties of the $W$ boson are independent of the cut-off scale, well behaved at high energies and they decouple for a large compactification scale. Their effects are dominant over those induced at the three--level by higher canonical dimension operators, consistently with the results reported in the literature.

\end{document}